\documentclass[twocolumn]{article}  
\usepackage[dvips]{graphicx}
\usepackage{amssymb,natbib}

\begin{document}
   \title{Introducing {\sc adaptsmooth}\thanks{Available at
    \texttt{http://www.wiki-site.com/index.php/ADAPTSMOOTH}}, a new
  code for the adaptive smoothing of astronomical images}

%   \subtitle{I. Overviewing the $\kappa$-mechanism}

\author{Stefano Zibetti\thanks{Current address: DARK Cosmology Centre,
    Niels Bohr Institute, University of Copenhagen, Juliane Maries Vej
    30, 2100 Copenhagen \O, Denmark,
    zibetti@dark-cosmology.dk}\\Max-Planck-Institut f\"ur Astronomie,
  K\"onigstuhl 17, D-69117 Heidelberg, Germany }
%\date{October 25, 2010}
\date{}
   \maketitle

% \abstract{}{}{}{}{} 
% 5 {} token are mandatory
 
   \abstract 
   {We introduce and publicly release a new standalone code, {\sc
     adaptsmooth}, which serves to smooth astronomical images in an
   adaptive fashion, in order to enhance the signal-to-noise ratio
   (S/N). The adaptive smoothing scheme allows to take full advantage
   of the spatially resolved photometric information contained in an
   image in that at any location the minimal smoothing is applied to
   reach the requested S/N. Support is given to match more images on
   the same smoothing length, such that proper estimates of local
   colours can be done, with a big potential impact on
   multi-wavelength studies of extended sources (galaxies,
   nebulae). Different modes to estimate local S/N are provided. In
   addition to classical arithmetic-mean averaging mode, the code can
   operate in median averaging mode, resulting in a significant
   enhancement of the final image quality and very accurate flux
   conservation. To this goal also other code options are implemented
   and discussed in this paper. Finally, we analyze in great detail
   the effect of the adaptive smoothing on galaxy photometry, in
   particular to check for conservation of surface brightness (SB) and
   aperture-integrated fluxes: deviations in SB with respect to the
   original image can be limited to $<0.01$~mag, with flux difference
   in apertures of less than 0.001 mag.  With respect to already
   existing adaptive smoothing codes, mainly developed by the X-ray
   community, {\sc adaptsmooth} uniquely provides the median averaging
   mode and a greater flexibility to adapt to different noise regimes
   and degrees of image contrast, which can occur especially in
   optical and near-IR imaging.}


\bibliographystyle{aa}
\section{Introduction}\label{intro_sec}
Extended astronomical sources, like galaxies and nebulae, are
characterized by a huge dynamical range of surface brightness (SB)
throughout their extent. This range spans several orders of magnitude
and can be captured by current astronomical imaging devices at optical
and near-IR wavelengths. However, as a consequence the signal-to-noise
ratio (S/N) of the information contained in the image varies wildly
depending on the (surface) brightness at any given position. Therefore
the problem arises how to extract the information by optimally
balancing the S/N of the flux measurement and the effective spatial
resolution.

Imaging devices are typically operated in background noise limited
conditions, i.e. the main source of noise comes from the background
rather than from the Poisson photon noise of the source itself, which
becomes dominant only at very high S/N levels. While the background
noise puts a well defined lower limit to the brightness of very
localized/point-like sources that can be detected at a given S/N, when
imaging diffuse objects the detection limit and the S/N at which a
given surface brightness is measured can be improved by combining the
flux measured in increasingly more pixels. Two approaches are
available to do this: binning and filtering (or smoothing). In the
first case, a number $N$ of pixels that share a similar intensity
level and/or match given spatial constraints are averaged, resulting
in an improvement in S/N of $\sqrt N$, in the ideal case of uniform
uncorrelated noise. Examples of the binning approach are azimuthally
averaged profiles of the surface brightness of galaxies along
elliptical isophotes \citep[e.g. the algorithm of][as implemented in
the {\sc iraf} task {\it ellipse}]{busko_96,jedre87} and the
tessellation algorithms that are extensively utilized in the analysis
of integral field spectroscopic observations \citep[for an
implementation see][]{cappellari_copin_2003}. As a result of binning,
the number of resolution elements is decreased (hence the spatial
resolution is worsened or even the dimensionality is lowered) and all
the available information in the image is compressed at a higher S/N
level. The filtering (or smoothing) approach consists in cutting the
highest spatial frequencies that contribute most of the noise. The
result of filtering is a smoothed image, where the intensity of each
pixel is replaced by a weighted average of the neighbouring pixels,
over an extent and with a weighting scheme that depend on the
particular kernel adopted. Popular choices of kernels are 2D gaussians
and top-hat functions (resulting in the so-called `boxcar'
smoothing). As with the binning, also in the filtering approach the
S/N is increased by a factor of the order of $\sqrt N$ (where $N$ is
the number of pixels in the kernel). Although the number of spatial
elements (pixels) is preserved, the information in the smoothed image
is correlated and spatial resolution is lost, proportionally to the
kernel size. In fact, by cutting the highest spatial frequencies not
only the noise but also the sharpest features of the image are thrown
away.

Both approaches have of course their own pros and
cons. Filtering/smoothing, in general, offers the best 2D rendition of
the information, although it is often non-optimal in terms of
compromise between spatial resolution and S/N enhancement, which,
reversely, is a point of strength of the tessellation algorithms. In
the regions of high surface brightness (SB) the S/N is high, such that
no or minimal smoothing is requested, while smoothing with
increasingly larger kernels is required to enhance the S/N of lower
and lower SB regions to acceptable levels. The optimal solution is to
apply to the image a smoothing with variable kernel size, which
matches the local S/N: this is what is called {\it adaptive
  smoothing}.

The X-ray community, pushed by the necessity to deal with diffuse
sources in very low S/N regimes, has produced a number of codes to
implement this basic concept. A (possibly incomplete) list includes:
{\sc fadapt} \citep[distributed as part of the package {\sc
  ftools},][]{blackburn_95}; \cite{huang_sarazin_96}; {\sc asmooth}
\citep[included in the Chandra data reduction package CIAO, also known
as {\sc csmooth} in the IDL implementation,][]{ebeling+06};
\cite{sanders_06}; {\sc asmooth} in the XMM-Newton Science Analysis
System.

The new, standalone, adaptive smoothing code {\sc adaptsmooth}
presented in this paper has been developed with in mind mainly typical
optical and near-IR imaging (although its application at other
wavelengths is well supported too) and allows to easily handle
different S/N regimes, even in cases where very limited information
about the noise properties of the images are available, as detailed in
Sec. \ref{subsec:noise}. We introduce {\it median} adaptive smoothing
in addition to the {\it mean} adaptive smoothing, which is the only
method implemented in the other codes mentioned above. The improvement
deriving from using the median as average estimate is especially
notable in presence of strong intensity gradients (e.g. bright
point-like sources overlaid on an area low surface brightness) and is
thoroughly illustrated in Sec. \ref{sec:testgal}. We also provide
different methods to prevent cross-talking between pixels at different
S/N levels, which can be switched on or off by the user (see
Sec. \ref{subsec:xtalk}) as opposed to the rigid implementation
\citep[e.g. {\sc a/csmooth},][]{ebeling+06} or lack of them in other
codes. Similarly to {\sc a/csmooth} in CIAO and {\sc asmooth} in the
XMM package, {\sc adaptsmooth} outputs maps of the smoothing length at
each location (``smoothing masks'') and allows to use them as input
for smoothing other images: this feature allows matching smoothing
scales across different images and is therefore essential to produce,
e.g., color maps (see Sec. \ref{subsec:inmask}).

{\sc adaptsmooth} is made publicly available via this URL
\texttt{http://www.wiki-site.com/index.php/ADAPTSMOOTH} or on
request to the author. In the following sections we describe the
concept of the algorithm and the actual code implementation with the
available options. We show a couple of examples on astronomical
images. Finally we analyze in detail the reliability of the code
output in terms of conservation of flux and surface brightness in
order to perform accurate surface photometry (specifically of
galaxies) in different conditions.

\section{Adaptive smoothing with {\sc adaptsmooth}}
\subsection{Concept}\label{subsec:concept}
We design the {\sc adaptsmooth} code to smooth an input image with a
variable-size kernel in order to provide the measure of the local
surface brightness with S/N equal to (or larger than) a minimum
user-provided value at any location. The size of the kernel is
determined locally (at each pixel) and iteratively as the minimum size
that allows to reach the requested S/N. In this sense, the {\sc
  adaptsmooth} algorithm is optimized to retain the maximum of spatial
resolution that is compatible with the requested S/N. As mentioned
above, this basic concept is common to most adaptive smoothing
codes. The main differences among codes reside in the way S/N is
defined and computed and in the way smoothing is actually performed
(type of kernel, weighting, FFT vs window sliding convolution, etc.).

Here is a description of basic steps of the {\sc adaptsmooth}
algorithm. We assume that a background-subtracted image and the
requested minimum $S/N$ are input. At each pixel $i$ an estimate of
the local noise $n_{i,1}$ is performed (see Section
\ref{subsec:noise}). The pixel brightness $f_{i,1}$ is compared with
the noise and if $f_{i,1}/n_{i,1} > S/N$ then the current pixel value
is retained and the procedure goes to the next pixel, else smoothing
is requested. In this case, a new estimate of the local surface
brightness $f_{i,2}$ is computed as the average in a circle of radius
1 pixel centered on the pixel $i$. The corresponding uncertainty
$n_{i,2}$ is computed and a new S/N check is performed: if
$f_{i,2}/n_{i,2} > S/N$ then the current SB estimate $f_{i,2}$ is
assigned to the pixel, else further smoothing is required. The
procedure is repeated increasing the radius of the circle by 1 pixel
each time and is iterated until the S/N condition is satisfied or the
maximum radius for the circle is reached. It must be noted that in the
ideal case of uncorrelated noise and no systematic background offsets,
the S/N condition is met in any case after a large enough number of
iterations, as $f_{i,l}/n_{i,l}\propto l$ (in the following we
call $l$ `smoothing level'). In reality, beyond a smoothing level of a
few tens correlated noise and systematics prevent the smoothing
procedure to produce any real S/N enhancement.

We note that {\sc adaptsmooth}, contrary to other codes, just
implements top-hat kernels, such that no (i.e. uniform) weighting is
applied when averaging. We do not feel this is a relevant limitation
since weighting more on the central pixels when the distribution is
dominated by noise is not expected to change the results
significantly. On the other hand, the speed of the algorithm can
substantially benefit from avoiding weights in computing average
quantities.

In {\sc adaptsmooth} the user can choose between median and arithmetic
mean as the average that is used to estimate the local SB inside the
kernel radius. This feature is unique to the present code. As we
demonstrate in Sec. \ref{sec:testgal}, the two methods perform very
similarly in absence of strong peaks in the SB distributions, but
using the median results in a much better behaved response in the
vicinity of strong discontinuities. However, one notable case where
mean averaging must be used in place of median averaging is when the
photon statistics is extremely poor, at the limit of binarity. This
occurs for example in UV or X-ray images where most of the pixels have
either one or zero counts: in this case one must use mean averaging in
order to obtain a fair representation of the surface brightness field,
while median averaging would produce just an almost random binary map
\citep[see][for a successful application of {\sc adaptsmooth} with
mean averaging in this regime]{salim_rich_10}.

In the next two sub-sections we describe the features of the code that
allow the user an optimal computation of the noise and to keep the
generation of artifacts under control. In Sec. \ref{sec:testgal} the
effect of these features on images and surface photometry of galaxies
will be shown and tested in detail.

\subsection{Noise estimates}\label{subsec:noise}
A realistic estimate of the local noise is central to properly
determine the kernel size for smoothing. {\sc adaptsmooth} offers
three possible methods to estimate noise.
\begin{itemize}
\item[\textit{i)}]{Poisson+background noise -- The following equation
    for the noise $\sigma$ is assumed:
    \begin{equation}\label{eq:1}
      \sigma^2=\sigma_{\mathrm{bkg}}^2+f/G  
    \end{equation}
    where $f$ is the brightness of the pixel in counts (analog-digital
    units, ADU), $\sigma_{\mathrm{bkg}}$ is the background
    r.m.s. noise in ADU (which includes the contributions from the sky
    background, readout noise and dark current) and $G$ is the gain,
    i.e. the number of photo-electrons per ADU. The last two numbers
    must be measured/known beforehand and input to the code, which
    assumes they are constant throughout the image. In the hypothesis
    of uncorrelated noise, when the smoothing circular aperture is
    considered, the background component of the noise
    $\sigma_{\mathrm{bkg}}$ is simply assumed to rescale as $\sqrt N$,
    $N$ being the number of pixels in the aperture. The
    Poisson+background noise is the recommended mode in all cases,
    unless strong variations of $\sigma_{\mathrm{bkg}}$ or $G$ are
    present in the image.}
\item[\textit{ii)}]{Background-dominated noise -- In most astronomical
    image applications, in the regime of SB where smoothing becomes
    relevant, one can safely assume background dominated noise, which
    corresponds to Eq. \ref{eq:1} with infinite gain or $f/G=0$. This
    is done in background-dominated noise mode, for which only
    $\sigma_{\mathrm{bkg}}$ is to be input. This can be particularly
    useful when the effective gain is not easily accessible or
    computable (e.g. near-IR stacks).  }
\item[\textit{iii)}]{Direct local noise estimate -- {\sc adaptsmooth}
    can compute the noise locally, without any prior knowledge of the
    image properties, as the r.m.s. counts in a circular aperture
    centered on the current pixel. A radius 2 pixels larger than the
    current smoothing level is adopted in this case to ensure
    sufficient statistics especially at the lowest smoothing
    levels. This mode offers a viable solution for images in which
    background dominated noise cannot be assumed yet the photon
    statistics is ill-determined, like, for instance, digitized
    photographic plates or digital images with unknown and/or
    significant gain or background noise variations across the
    field. However, modes \textit{i)} and \textit{ii)} should be
    preferred whenever possible, in order to avoid the artifacts that
    can arise from adopting direct local noise calculation. Especially
    in the proximity of sharp features the local r.m.s. is dominated
    by SB fluctuations due to the real objects, which leads to severe
    noise overestimation and hence oversmoothing.}
\end{itemize}

To our knowledge, {\sc adaptsmooth} is the only code offering these
three choices of noise computation\footnote{{\sc fadapt} just
  considers pure Poisson noise, CIAO {\sc csmooth} supports background
  dominated and Poisson noise, and XMM {\sc asmooth} support only
  Poisson mode and user pre-computed variance maps.}, which cover all
possible user's requests in typical optical/near-IR imaging.

\subsection{Cross-talking between smoothing levels}\label{subsec:xtalk}
A potential problem of the adaptive smoothing scheme is that the
estimated SB of pixels in high smoothing levels (i.e. low original
S/N) is affected by neighbouring pixels at lower smoothing levels,
which have higher S/N and brightness: we call this effect
`cross-talking' between smoothing levels. As we show in
Sec. \ref{sec:testgal}, the cross-talking can artificially broaden
sharp bright features (such as stars), especially when the arithmetic
mean is chosen as average estimator instead of the median. Moreover,
this effect may result in not conserving the mean SB of the image,
because of the asymmetry of the cross-talking: bright pixels require
less smoothing and therefore are only marginally affected by
neighbouring lower-SB pixels, while lower-SB pixels may be made
significantly brighter by smoothing over larger areas that might
include bright pixels.

To limit the cross-talking we introduce the option of excluding pixels
with lower smoothing levels from the computation of the average
smoothed SB. Specifically, the `level cut' parameter `c' can be
specified to exclude either smoothing level 1 from the other levels
(c$=1$), or to make a relative cut to exclude all levels but those
higher than the current $-n$ (any c$<0$, c$=-n$). This option is
switched off by default.

We note that CIAO {\sc csmooth} implements a similar solution to limit
cross-talking, which essentially corresponds to setting c$=-1$ in {\sc
  adaptsmooth}. However, contrary to {\sc adaptsmooth} which allows
the user full flexibility, this option is hard-coded in {\sc csmooth}
and no choice is given to the user if and to which degree to use it.

\subsection{Matching images in input-mask mode}\label{subsec:inmask}
By default {\sc adaptsmooth} determines the smoothing level of each
pixel by a direct analysis of the image, as explained above, and
outputs the smoothed image and an image containing the smoothing level
of each pixel, which we dub `smoothing mask'. This is useful in first
place for checking purposes, to understand how the image has been
elaborated. Many astronomical applications make use of multi-band
imaging to derive spatially resolved physical information, for
instance colour maps that serve to determine the stellar mass
distribution or the spatial variation of the SED in a galaxy,
flux-ratios maps in narrow-bands to extract extinction maps in
star-forming regions, etc. For all this kind of applications it is
crucial that the spatial resolution of the images in different bands
are perfectly matched, so that the flux measurements are
consistent. {\sc adaptsmooth}-ed images have varying effective
resolution across the field depending on the S/N of the pixels, which
in turn depend on the observed band and the observing conditions. In
order to obtain consistent flux measurements in two or more bands it
is necessary to smooth the corresponding images in the same way. To do
this we allow the user to force {\sc adaptsmooth} to use
pre-determined smoothing levels.

The standard procedure to match a number of {\sc adaptsmooth}-ed
images (which we assume to be already accurately registered to the
same plate scale) is as follows. First, one has to run {\sc
  adaptsmooth} on the individual images in the standard mode to obtain
the requested S/N (which can be different in each image); for each
image, the smoothing mask is output, which contains the {\it minimum}
smoothing level required at each pixel to obtain the requested
S/N. The smoothing masks have to be combined to retain the maximum
value at each pixel. Finally, {\sc adaptsmooth} needs to be run a
second time on each image by giving the combined smoothing mask {\it
  as input}. In this way the requested minimum S/N is ensured at all
pixels, while the same smoothing is performed in all images.

\subsection{Technical notes on the code}\label{subsec:code}
{\sc adaptsmooth} is a standalone code written in C language and works
as line command with standard GNU options parsing. Graphical user
interface will be provided in future releases. All image input/output
is based on the standard C library CFITSIO \citep{cfitsio}, not
provided with the code.

Execution time depends critically on the chosen mode, the properties
of the image and the requested S/N. In our tests with a 2.4 GHz
processor typical execution times range between a few seconds up to
5-6 minutes for a 1k$\times$1k image. Mask-input mode is always faster
than the standard mode; using mean average is slightly faster than
median average; local-noise takes significantly longer than Poisson or
background only; with `level cut' enabled the execution time is
practically twice as much as in standard mode with the same set of
parameters.

In order to speed up the access to the pixels in the smoothing
apertures, their relative coordinate offsets with respect to the
current image position are pre-computed up to the 27th smoothing level
and stored in a static array. This limits the smoothing level that can
be used to 27.

In local noise mode {\sc adaptsmooth} approximates the r.m.s. with
half of the $16^{\mathrm{th}}-84^{\mathrm{th}}$ percentile range, as
it is more robust against outliers. For this and for median estimates
a fast selection algorithm is needed. After some testing, we opted for
an implementation of the algorithm by \cite{floyd_rivest_75}.

It is worth noting that when the `level cut' option is enabled (see
Sec. \ref{subsec:xtalk}) the code works in two passes, hence it is
roughly twice as slow. The first pass provides a preliminary smoothing
level for each pixel and the second pass computes the actual smoothed
image where the level cuts are properly accounted for. Although in
principle more passes might be necessary to converge to the final
determination of smoothing levels, we checked that in most practical
cases two passes are enough.

\subsection{Enhancing imaging and photometry with {\sc
    adaptsmooth}}\label{subsec:example}
The adaptive smoothing technique may have a big impact on a number of
astronomical fields where spatially resolved flux information is
required over a large dynamical range of surface brightness.
\cite{zibetti_charlot_rix_09a,zibetti_charlot_rix_10} showed how the
adaptive smoothing can be applied to galaxies to map their colours and
eventually derive the stellar mass density distribution and its link
with local SED. \cite{mentuch+10} successfully exploited the power of
the adaptive smoothing technique to identify the regions responsible
for NIR excess in the SED of star-forming galaxies.  {\sc adaptsmooth}
has also been used by \cite{martinezdelgado+10} and Mart\'inez-Delgado
et al. (in preparation) to estimate optical colors and calibrate the
photometry of ultra-faint diffuse streams of stars around nearby
galaxies.

%%%%%%%%%%%%%%%%%%%%%%%%%%%%%%%%%%%%%%%%%%%%%%%%%%%%%%%%%%%%%%%%%%%%
\begin{figure*}
\includegraphics[width=\textwidth]{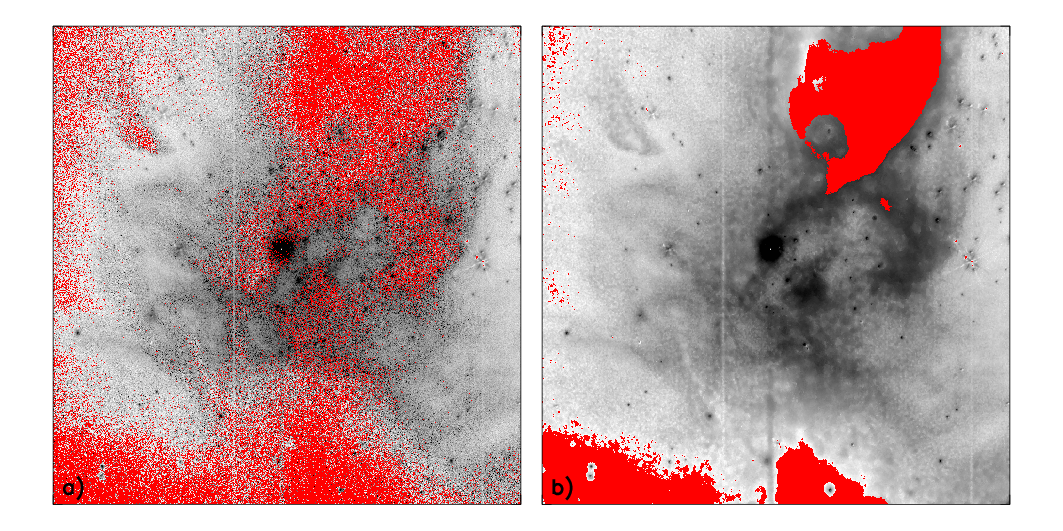}
\caption{The map of
  $\log~\frac{\mathrm{Paschen}\beta}{\mathrm{Brackett}\gamma}$ for the
  emission nebula NGC\,2024 \citep[from][]{bik+03}. Panel \textit{a)}
  shows the map as obtained from the original images. Panel
  \textit{b)} displays the map obtained after median-average smoothing
  the images with {\sc adaptsmooth} to a minimum S/N of 10 (background
  dominated noise is assumed). Red is used in panel \textit{a)} for
  pixels where fluxes are zero or negative in the original images; in
  panel \textit{b)} red identifies areas where the minimum S/N could
  not be reached. {\sc adaptsmooth} retains the highest resolution in
  the high-SB regions, while enhancing the S/N and therefore allowing
  meaningful measurements at low-SB by using the minimal required
  smoothing.}\label{fig:SFregion}
\end{figure*}
%%%%%%%%%%%%%%%%%%%%%%%%%%%%%%%%%%%%%%%%%%%%%%%%%%%%%%%%%%%%%%%%%%%%
In Fig. \ref{fig:SFregion} we illustrate the full power of the
adaptive smoothing technique on a emission-line ratio map of the
star-forming H{\sc ii} region NGC\,2024
\citep{bik+03}. $\log~\frac{\mathrm{Paschen}\beta}{\mathrm{Brackett}\gamma}$
is computed for each pixel and shown as grey-scale in the two panels
of Fig. \ref{fig:SFregion}. Panel \textit{a)} (to the left) shows the
line-ratio map as obtained from the original images. The pixels
highlighted in red are those for which either of the two images has
zero or negative intensity. The right-hand panel \textit{b)} displays
the map obtained from {\sc adaptsmooth}ed images with minimum S/N of
10 in both bands, obtained using median averaging and assuming
background dominated noise. The smoothing in the two original
pass-band images is matched as explained in
Sec. \ref{subsec:inmask}. Pixels for which the minimum S/N could not
be reached are painted in red. It is immediate to see \textit{i)} the
large number of pixels for which the line ratio becomes measurable
only after {\sc adaptsmooth}ing and \textit{ii)} the detail which is
gained in the map thanks to the increased S/N provided by {\sc
  adaptsmooth}, which in this case guarentees errors smaller than 15
per cent on the line ratio in each individual pixel as opposed to the
huge scatter obtained without adaptive smoothing. Similarly to this
application, \cite{pasquali+11} use {\sc adaptsmooth} to
produce deep, high-resolution maps of near-IR line emissions in the
nearby dwarf star-bursting galaxy NGC\,1569, which in turn are used to
derive maps of dust extinction and exctinction-corrected
star-formation rate of unprecedented accuracy and detail.

This shows that by adaptively smoothing optical and near-IR images it
is possible to produce extremely accurate, deep and high-resolution
colour and line ratio maps using currently available datasets, at no
extra expense in terms of telescope time. In turn this opens up the
opportunity to perform spatially resolved analysis of SED and dust
extinction throughout extended object such as nebulae and galaxies.

\section{Testing {\sc adaptsmooth} with galaxy
  photometry}\label{sec:testgal}
In this section we test the performance of {\sc adaptsmooth} on the
image of a galaxy, using different modes and parameters. We choose
NGC\,5713 as imaged in $r$ band in the Sloan Digital Sky Survey
\citep[SDSS, ][]{SDSS}. This galaxy displays a complex enough
structure so that the code can be tested in relatively extreme
conditions, while the SDSS is chosen as representative of present-day
sources of astronomical images.
\subsection{Morphological tests}
%%%%%%%%%%%%%%%%%%%%%%%%%%%%%%%%%%%%%%%%%%%%%%%%%%%%%%%%%%%%%%%%%%%%
\begin{figure*}
\includegraphics[width=\textwidth]{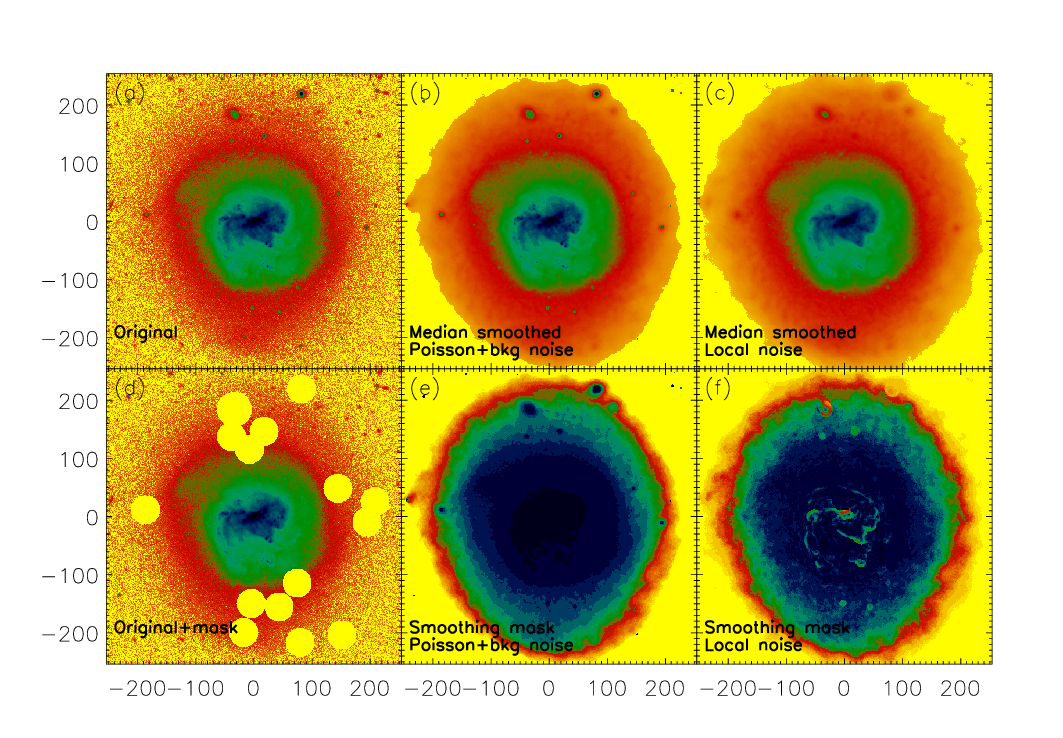}
\caption{The galaxy NGC\,5713 as our benchmark for testing {\sc
    adaptsmooth}. Original SDSS $r$-band image (panel \textit{a)});
  {\sc adaptsmooth}-ed image (median-averaging, S/N$>20$) with
  Poisson+background noise estimation (panel \textit{b)}); {\sc
    adaptsmooth}-ed image (median-averaging, S/N$>20$) with local
  noise estimation (panel \textit{c)}); original SDSS $r$-band image
  with overlaid masks for problematic regions (panel \textit{d)});
  smoothing mask relative to panel \textit{b)} (panel \textit{e)};
  smoothing mask relative to panel \textit{c)} (panel \textit{f)}. The
  colour scales for the images cover the entire dynamical range with a
  logarithmic stretch, which is kept the same in all images. For the
  masks, colours go linearly from dark blue (smoothing level 1) to
  orange/yellow (level 20).}\label{fig:galaxy_adsmooth}
\end{figure*}
%%%%%%%%%%%%%%%%%%%%%%%%%%%%%%%%%%%%%%%%%%%%%%%%%%%%%%%%%%%%%%%%%%%%
The panel \textit{a)} of Fig. \ref{fig:galaxy_adsmooth} shows the
original SDSS $r$-band image of the galaxy. Panels \textit{b)} and
\textit{c)} show the {\sc adaptsmooth}-ed images obtained for a
minimum S/N of 20, with median average, maximum smoothing level of 20
and no level cuts. In panel \textit{b)} the Poisson+background noise
estimate is used, whereas a purely local noise estimate from the image
is adopted in panel \textit{c)}. A careful comparison of the two
smoothed images shows that a number of localized sharp features that
are seen in \textit{b)} completely disappear, smoothed out, in
\textit{c)}. These include stars, but also star-forming regions and
bright stellar knots in the central regions of the galaxy. A
comparison of the corresponding smoothing masks (panels \textit{e)}
and \textit{f)}, where the smoothing levels are coded from blue to
red/yellow going from 1 to 20) highlights the fact that noise
estimates based on the local r.m.s. result in grossly overestimating
the smoothing radii requested in the vicinity of sharp features. While
the Poisson+background noise model correctly recognizes that nowhere
in the central parts of the galaxy substantial smoothing is needed to
enhance the S/N, based on the local r.m.s. the code performs massive
smoothing at several locations down into the nucleus. On the other
hand, far from sharp structures the smoothing levels assigned with the
two methods are in excellent agreement, as one can see looking at the
overall structure of the smoothed images and, even better, of the
smoothing masks. To summarize, the local noise mode can give
substantially correct results in smooth areas, but should be avoided
in presence of sharp structures.

We have also compared the Poisson+background noise mode with the pure
background noise mode (not shown) and find practically no difference,
with the exception of a very small number of pixels jumping from
smoothing level 2 to 1 (i.e. no smoothing) when the Poisson
contribution is neglected. This was expected as we know that SDSS
images are in the regime of background dominated noise. In the rest of
the analysis we consider only noise estimated based on the
Poisson+background model.

\begin{figure*}
\includegraphics[width=\textwidth]{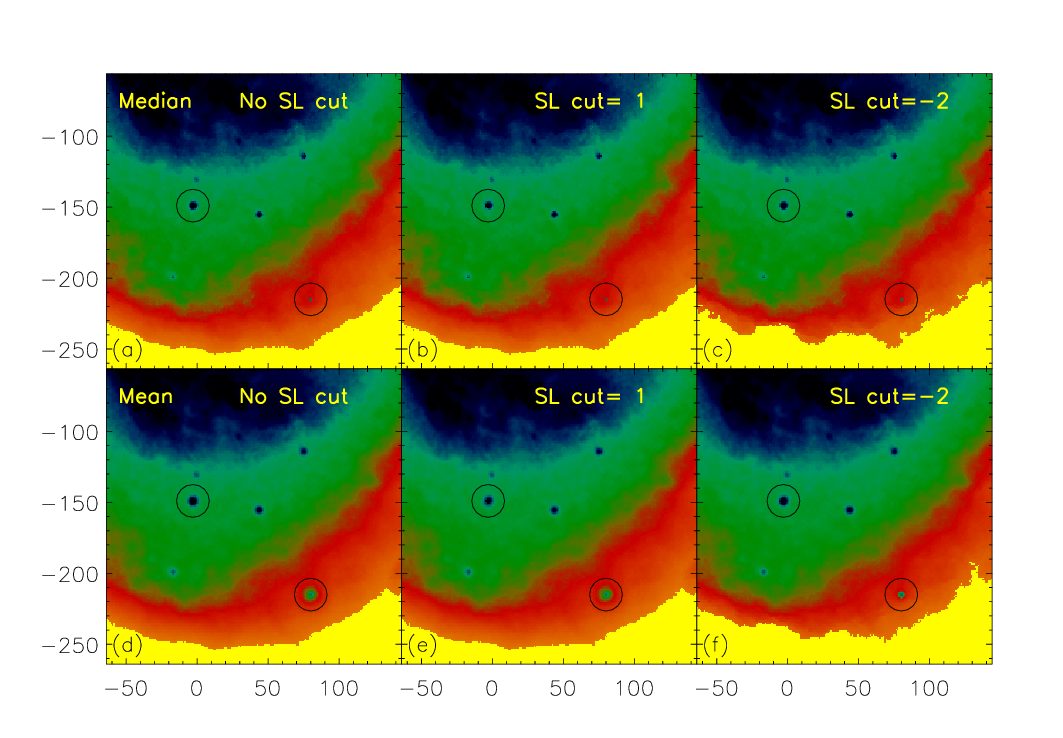}
\caption{The effect of different smoothing modes and smoothing level
  cuts are illustrated in this Figure, for a zoomed-in region of
  NGC\,5713. The \textit{top} row (panels \textit{a), b), c)}) show
  images obtained using median-averaging, while for the
  \textit{bottom} row (panels \textit{d), e), f)}) mean-averaging is
  adopted. In the \textit{first} column (panels \textit{a), d)}) no
  smoothing level cut is used; in the images of the \textit{second}
  column (panels \textit{b), e)}) level 1 is excluded from the
  computation of the other levels; the \textit{third} column (panels
  \textit{c), f)}) shows the results for excluding all levels up to
  the current $-2$. Two features are highlighted (black circles) where
  the effects of the different smoothing modes are particularly
  evident.}\label{fig:med_mean_cuts}
\end{figure*}

In Fig. \ref{fig:med_mean_cuts} we compare smoothed images obtained
with the two different average options, median (upper row) and mean
(lower row), and with different level cuts. In this figure we show
only a zoomed-in region of the same galaxy NGC\,5713, in order to
better highlight the differences. The first panel of each row
(viz. \textit{a)} and \textit{d)}) shows the results for no smoothing
level cuts, the second panel (viz. \textit{b)} and \textit{e)}) is
obtained excluding level 1 from all other levels, and the third panel
(viz. \textit{c)} and \textit{f)}) is the image obtained by excluding
all smoothing levels up to the current one minus 2.
%%%%%%%%%%%%%%%%%%%%%%%%%%%%%%%%%%%%%%%%%%%%%%%%%%%%%%%%%%%%%%%%%%%%
%%%%%%%%%%%%%%%%%%%%%%%%%%%%%%%%%%%%%%%%%%%%%%%%%%%%%%%%%%%%%%%%%%%%
The first thing to note in Fig. \ref{fig:med_mean_cuts} is that using
the mean results in broadening the SB peaks over significantly large
areas, especially when the peak is in the middle of a low-SB region
where substantial smoothing is applied. This is particularly evident
for the source located at (80,-215), but also for other sources, most
likely foreground stars. The median smoothing is more robust against
this cross-talking effect because a few high-SB pixels falling inside
a large smoothing aperture, where most pixels have much lower SB,
negligibly affect the median value, but can substantially alter the
mean.  Smoothing level cuts are meant to reduce the cross-talking
effect. From panels \textit{a)}, \textit{b)} and \textit{c)} we see
that the cuts do not produce any visible effect on the broadening of
bright peaks when the median is used, confirming that the median
estimate already produces optimal results in this sense. On the
contrary, using smoothing level cuts when the mean average is adopted
can strongly reduce the broadening (panels \textit{d)}, \textit{e)},
\textit{f)}). Excluding only the highest-SB pixels (level 1) can help
only with very bright sources, such as the star at (0,-148), while a
relative cut (in this case, considering only the levels higher than
the current one minus 2) significantly reduces the broadening for all
peaks. A second effect of using a relative cut is seen at the lowest
SB, where the exclusion of higher SB levels produces lower SB in the
images with cuts with respect to those without cuts. This effect can
be seen comparing the yellow area, which corresponds to minimum S/N
not reachable, of panels \textit{c)} (or \textit{f)}) and \textit{a)}
(or \textit{d)}): applying no cuts results in a larger area with
acceptable S/N.

\subsection{Quantitative tests}
\begin{figure*}
\includegraphics[width=\textwidth]{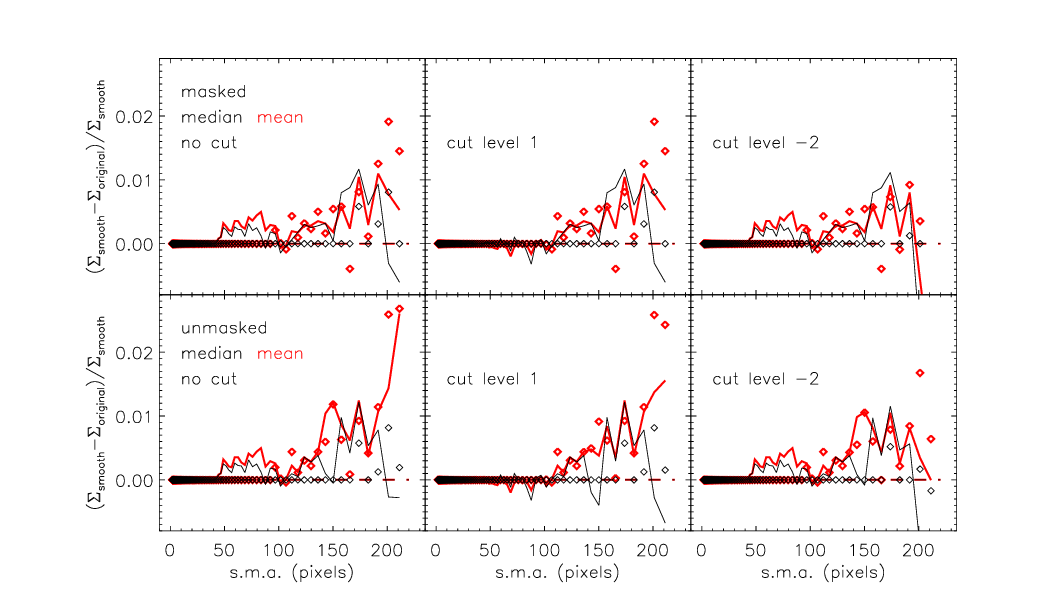}
\caption{Statistics of relative difference of pixel brightness in
  NGC\,5713 after and before {\sc adaptsmooth}-ing, as a function of
  radius and for different modes and parameters (as indicated in the
  legends, black is for median-smoothing, red for
  mean-smoothing). Solid lines show the mean relative differences,
  while diamonds are median differences. The \textit{bottom} row is
  relative to the full image, in the \textit{top} row masked areas
  (see Fig. \ref{fig:galaxy_adsmooth}) are
  neglected.}\label{fig:pixbypix_diff}
\end{figure*}

\begin{figure*}
\includegraphics[height=0.6\textheight]{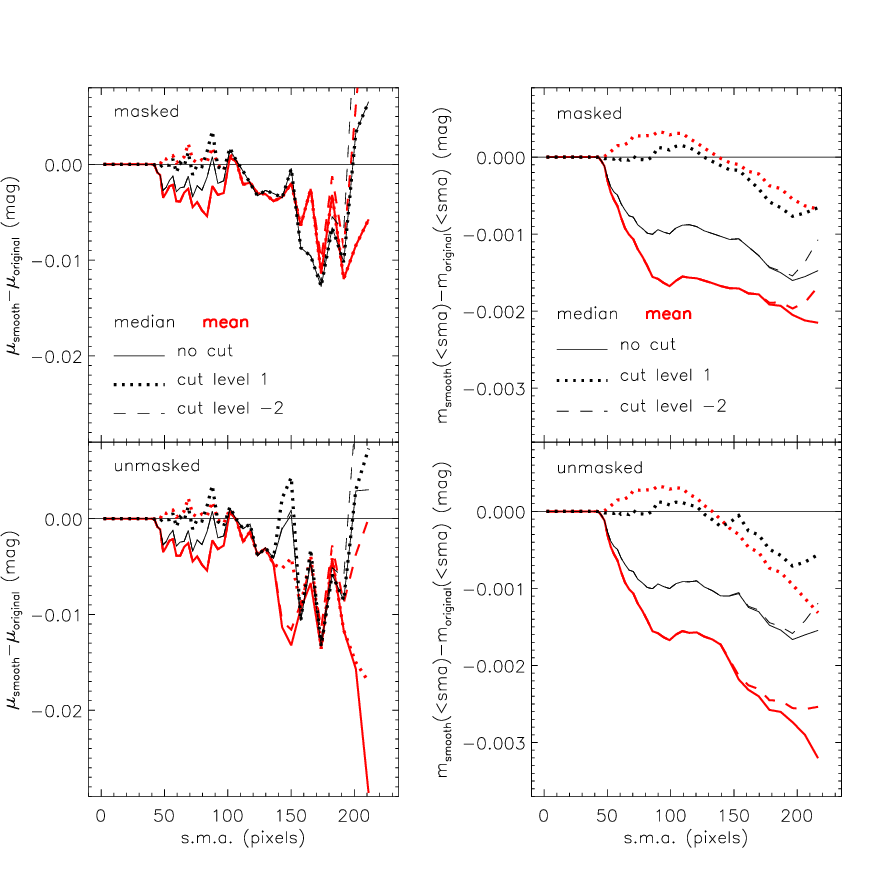}
\caption{Deviations of the SB-profiles (\textit{left} column) and of
  the growth curves (\textit{right} column) of {\sc adaptsmooth}-ed
  images of NGC\,5713 with respect to the original. As indicated in
  the legends, black is for median-smoothing, red for mean-smoothing;
  different lines are used for different smoothing level cuts. The
  \textit{bottom} plots refer to the full image, in the \textit{top}
  plots masked areas (see Fig. \ref{fig:galaxy_adsmooth}) are
  neglected. }\label{fig:prof_diff}
\end{figure*}

In Figures \ref{fig:pixbypix_diff} and \ref{fig:prof_diff} we analyze
how the adaptive smoothing with different modes and parameters affects
the SB distributions in pixels, on a pixel-by-pixel base and in terms
of azimuthally averaged profiles and aperture photometry,
respectively.  The analysis performed in this section is not meant to
be illustrative of typical possible applications of {\sc adaptsmooth}
to galaxy photometry. In fact, all photometric measurements involving
binning, averaging and integrating within apertures are better
performed on original unsmoothed images. Here we just want to show
that the output of {\sc adaptsmooth} is not biased and average surface
brightness and fluxes in different apertures are conserved.

In first place we bin all pixels in the image in elliptical annuli,
centered on the galaxy centre and with ellipticity and position angle
corresponding to the outer isophotes. Secondly, we consider two
versions of the images: the full image and one where some regions are
masked out (shown in panel \textit{d)} of
Fig. \ref{fig:galaxy_adsmooth}). The masked regions correspond to
bright spots on top of a smoother low-SB background, where the largest
differences between different methods is observed, as shown in the
previous paragraph.
%%%%%%%%%%%%%%%%%%%%%%%%%%%%%%%%%%%%%%%%%%%%%%%%%%%%%%%%%%%%%%%%%%%%
%%%%%%%%%%%%%%%%%%%%%%%%%%%%%%%%%%%%%%%%%%%%%%%%%%%%%%%%%%%%%%%%%%%%

In the six panels of Fig. \ref{fig:pixbypix_diff}, we consider
\textit{for each pixel} the relative difference between the SB in the
smoothed image and the original image, given by
$\delta=(\Sigma_{\mathrm{smooth}}-\Sigma_{\mathrm{original}})/\Sigma_{\mathrm{smooth}}$.
With open diamonds we plot the median value of $\delta$ in each
elliptical annulus, as a function of the semi-major axis: this
quantity illustrates the balance between the number of pixels that are
scattered above and below their original value. The solid lines
represent the mean value of $\delta$, which gives a measure of how
well the SB is conserved. Black symbols and lines are used for the
median smoothing mode, while the red ones are for the mean
smoothing. As in Fig. \ref{fig:med_mean_cuts}, the three panels in
each row are for no smoothing level cut, cut level 1, and the cut to
exclude all but the levels above the current minus 2. The two rows
show the results for the full image (bottom row) and for the image
where masked regions have been excluded (top row).  The main thing to
note is that, with very few exceptions, all deviations are positive:
the adaptively smoothed images have a bias to higher surface
brightness with respect to the original. The reason for this bias is
essential to the adaptive scheme: as the smoothing aperture is
increased until the requested S/N is reached, the code tends to get
rid of negative noise fluctuations by forcing a stronger smoothing,
while positive fluctuations are not contrasted. This bias is of the
order of 2\% at most if the whole image is considered, 1\% if we
neglect the bright spots (masked regions).

Let's now consider the smoothing modes without level cuts (panels in
the first column of Fig. \ref{fig:pixbypix_diff}) in more detail. By
using {\it median} smoothing, the median $\delta$ is 0 with only a few
exceptions at large radii (i.e. low SB) and this indicates that the
number of pixels with positive and negative variations almost balance,
as expected for a symmetric noise distribution. However, the mean
$\delta$ has most of the time positive values, as a consequence of the
bias discussed before. It is noticeable that the masked regions have
relatively little impact both on the median and on the mean bias, thus
indicating the robustness of the median smoothing mode. On the
contrary, the {\it mean}-smoothed images are generally more strongly
biased toward higher SB, as shown by both median and mean $\delta$
profiles. Also, by comparing the top and bottom panels, and as already
seen in the previous section, the mean smoothing is much more
sensitive to and, hence, biased by, the presence of bright spots on
top of low-SB regions.

The top row of Fig. \ref{fig:pixbypix_diff} demonstrates that, once
the problematic regions are taken out, the use of level cuts either in
mean- or median-smoothing mode makes essentially no difference. The
only notable exception is that cutting level 1 reduces the relative SB
bias from 0.003 to $<0.001$ in the region between 50 and 100 pixels,
which corresponds to the transition from the high-SB where no
smoothing is required to lower-SB. On the other hand, the bottom row
of Fig. \ref{fig:pixbypix_diff} demonstrates that on the full image
(including the masked regions, which is the normal case in practice)
the use of level cuts can significantly reduce the SB bias below 1\%
level in mean-smoothing mode, hence comparable to what one gets in
median-smoothing mode.

%%%%%%%%%%%%%%%%%%%%%%%%%%%%%%%%%%%%%%%%%%%%%%%%%%%%%%%%%%%%%%%%%%%%
%%%%%%%%%%%%%%%%%%%%%%%%%%%%%%%%%%%%%%%%%%%%%%%%%%%%%%%%%%%%%%%%%%%%
In Fig. \ref{fig:prof_diff} we show how the smoothing performed with
{\sc adaptsmooth} affects two classical photometric measurements, that
is azimuthally averaged SB profiles and growth curves of magnitude. In
the left-hand panels we plot the difference in surface brightness
between the azimuthally averaged profiles obtained on the {\sc
  adaptsmooth}-ed images with different modes and parameters and the
original image, both for the masked and the full images (upper and
lower panels, respectively). These plots basically mirror the mean
curves of Fig. \ref{fig:pixbypix_diff} and show that the SB in
smoothed images is biased brighter by no more than 0.01 mag
approximately. The same remarks about smoothing modes and level cuts
as made for Fig. \ref{fig:pixbypix_diff} apply also here.

The growth curves (right-hand panels) show that flux in conserved in
the smoothed images at better than 0.003 mag (0.002 mag when
problematic regions are masked out). The regions that have the largest
influence in biasing the total flux high are not the bright spots on
top of low-SB regions, however, but those regions at the transition
between no smoothing and smoothing level 2. Using median- instead of
mean-smoothing already reduces this effect by $\approx 50$\%, although
the best results are obtained by cutting level 1 in addition: in this
case the flux is conserved with an accuracy of $<0.001$~mag. Other
level cuts that do not exclude level 1 have marginally influence on
flux conservation.

As a general conclusion of this section on testing {\sc adaptsmooth}
on galaxy photometry, we can say that the best results in terms of
minimizing artifacts, conserving surface brightness and total flux are
obtained with median-smoothing mode and by cutting smoothing level 1.
We remind the reader and the user that this configuration may not be
ideal in all cases, depending on the typical contrast of the images to
be processed and their photon statistics. A notable example is the
case of extremely low photon-count rates, discussed at the end of
Sec. \ref{subsec:concept}, where mean-smoothing has definitely to be
preferred to median-smoothing. {\sc adaptsmooth} offers the user full
flexibility and a full range of options (which are unavailable or only
partly available in other public codes) which can be best adapted to
different imaging datasets.

\section{Summary}
We have introduced {\sc adaptsmooth}, a new standalone code to perform
adaptive smoothing of astronomical images, which is made publicly
available. {\sc adaptsmooth} uses either median or arithmetic-mean
average within apertures whose size is variable and determined in
order to provide pixel surface brightness measurements above the
minimum requested S/N. In this way the best compromise between
photometric accuracy and spatial resolution is obtained at any
location of an image. The use of smoothing masks allows to work with
multi-band imaging and hence to spatially resolve the spectral energy
distribution of extended objects. Adaptive smoothing extends to much
dimmer surface brightness the ability of measuring local fluxes and
colours and hence can have a very strong impact on the study of
extended objects, such as galaxies \citep[as shown, e.g.,
in][]{zibetti_charlot_rix_09a,zibetti_charlot_rix_10} and nebulae (see
Fig. \ref{fig:SFregion}).

We provide and discuss different ways to estimate local S/N in an
image (based on Poisson+background noise, on background noise only, or
on local signal fluctuations) and options to reduce artifacts
(smoothing level cuts). Finally, we analyze in detail the qualitative
and quantitative effects of adaptive smoothing on galaxy images. In
the best operation mode for a typical optical image (i.e. median
smoothing and smoothing level cut 1) the azimuthally averaged SB
profile of the smoothed images deviates by less than 0.01 mag from
that of the original image, while the flux growth curve deviates by
less than 0.001 mag.

Compared to other existing codes for adaptive smoothing, {\sc
  adaptsmooth} provides the broadest variety of options to operate in
different regimes. In particular, {\sc adaptsmooth} uniquely
implements a median-averaging mode, which turns out to provide best
results in terms of flux conservation and minimization of artifacts
originating from bright sources when the photon statistics are well
populated. The use of smoothing level cuts is much more flexible than
in other codes (e.g. in {\sc csmooth}). Finally, the three possible
choices for S/N estimate cover essentially all regimes for typical
optical and near-IR images (as well as for shorter wavelengths) and
guarentee good results even with minimal knowledge of the data error
model. Such support is only partly provided by other adaptive
smoothing codes (e.g. {\sc csmooth} supports Poisson and background
dominated noise, but no direct r.m.s. estimate on the image, while XMM
{\sc asmooth} only supports Poisson errors or user provided variance
maps).

%%%%%%%%%%%%%%%%%%%%%%%%%%%%%%%%%%%%%%%%%%%%%%%%%%%%%%%%%%%%%%%%%%%%%%%%%%%%%
%
\section*{Acknowledgments}
We thank Paolo Franzetti and Anna Gallazzi for useful comments and for
helping in testing the code, and Arjan Bik for kindly providing the
images of NGC2024.
\bibliography{adaptsmooth}
\label{lastpage}
\end{document}